\documentclass[prd,aps,amsmath,amssymb,nofootinbib,twocolumn,preprintnumbers]{revtex4}
\pdfoutput=1

\usepackage{graphicx}
\usepackage{dcolumn}
\usepackage{hyperref}
\usepackage{bm}
\usepackage{amsmath}
\usepackage{amsthm}
\usepackage{amsfonts}
\usepackage{ifthen}
\usepackage{psfrag}
\usepackage{slashed}
\usepackage{hyperref}
\usepackage{color}

\def\ls{\mathrel{\lower4pt\vbox{\lineskip=0pt\baselineskip=0pt
           \hbox{$<$}\hbox{$\sim$}}}}
\def\gs{\mathrel{\lower4pt\vbox{\lineskip=0pt\baselineskip=0pt
           \hbox{$>$}\hbox{$\sim$}}}}
\def\drawbox#1#2{\hrule height#2pt
\hbox{\vrule width#2pt height#1pt \kern#1pt
              \vrule width#2pt}
              \hrule height#2pt}

\def\Asym#1#2{\vcenter{\vbox{\drawbox{#1}{#2}
              \kern-#2pt       % line up boxes
              \drawbox{#1}{#2}}}}

\def\nn{\nonumber}

\newcommand{\be}{\begin{equation}}
\newcommand{\ee}{\end{equation}}
\newcommand{\bea}{\begin{eqnarray}}
\newcommand{\eea}{\end{eqnarray}}

\newcommand{\gsim}{\lower.7ex\hbox{$\;\stackrel{\textstyle>}{\sim}\;$}}
\newcommand{\lsim}{\lower.7ex\hbox{$\;\stackrel{\textstyle<}{\sim}\;$}}

\newcommand{\met}{{E\!\!\!\!/_{T}}}

\newcommand{\ben}{\begin{enumerate}}
\newcommand{\een}{\end{enumerate}}
\newcommand{\bei}{\begin{itemize}}
\newcommand{\eei}{\end{itemize}}

\begin{document}

\preprint{MI-TH-1523}
\preprint{CETUP2015-011}

\title{Distinguishing Standard Model Extensions using Monotop Chirality at the LHC}

\author{Rouzbeh Allahverdi$^{1}$}
\author{Mykhailo Dalchenko$^{2}$}
\author{Bhaskar Dutta$^{2}$}
\author{Andr\'es Fl\'orez$^{3}$}
\author{Yu Gao$^{2}$}
\author{Teruki Kamon$^{2,4}$}
\author{Nikolay Kolev$^{5}$}
\author{Ryan Mueller$^{2}$}
\author{Manuel Segura$^{3}$}
\affiliation{
$^{1}$~Department of Physics and Astronomy, University of New Mexico, Albuquerque, NM 87131, USA \\
$^{2}$~Department of Physics and Astronomy,\\ Mitchell Institute for Fundamental Physics and Astronomy, Texas A\&M University, College Station, TX 77843-4242, USA\\
$^{3}$~Departamento de F\'isica,\\ Universidad de los Andes,\\ Bogot\'a, Carrera 1 18A-10, Bloque IP, Colombia\\
$^{4}$~Department of Physics, Kyungpook National University, Daegu 702-701, South Korea\\
$^{5}$~Department of Physics,\\ University of Regina,\\ SK, S4S 0A2, Canada\\
}

\begin{abstract}
We present two minimal extensions of the standard model, each giving rise to baryogenesis.
They include heavy color-triplet scalars interacting with a light Majorana fermion that can be the dark
matter (DM) candidate. The electroweak charges of the new scalars govern their couplings to quarks
of different chirality, which leads to different collider signals. These models predict monotop events
at the LHC and the energy spectrum of decay products of highly polarized top quarks can be used
to establish the chiral nature of the interactions involving the heavy scalars and the DM.
Detailed simulation of signal and standard model background events is performed,
showing that top quark chirality can be distinguished in hadronic and leptonic decays of the top quarks.
\end{abstract}

\maketitle

\section{Introduction}

The monojet final states have attracted attention as of late since they can probe many extensions of the standard model (SM) that include dark matter (DM) candidates~\cite{monojet}. Along the same line, the monotop final state is also under investigation as a probe of DM models~\cite{monotops}. 

Recently, a minimal extension to the SM has been proposed~\cite{Allahverdi:2013mza} that explains the proximity of baryon and DM abundances~\cite{cladogenesis} by introducing baryon number violating interactions via (a set of) heavy color-triplet scalars $X_\alpha$
and a light singlet Majorana fermion. 
The fermion becomes stable, and hence a viable DM candidate, when its mass is almost equal to the proton mass. 
Since no new discrete symmetry is needed to protect the DM particle against decay, 
this model naturally predicts monojet or monotop signals at the LHC with a characteristic resonance~\cite{Dutta:2014kia} that features a Jacobian peak in the jet's transverse momentum distribution with a sizable missing transverse energy ($\met$).
The model in addition produces dijet, dijet + $\met$, 4-jet + $\met$ final states.

For successful baryogenesis, the TeV $X$ fields can couple to quarks of any generation and chirality. While largely leaving the same imprint on early universe, the couplings to different generations can potentially lead to very different signals at the LHC. The first-generation quark couplings enhance the production rate of $X$, while the third-generation coupling yields a monotop final state with a sizable $\met$. Moreover, when monotop events are present, the top quark polarization can be a useful probe of the chiral property of the new interactions in the model. In this paper, we discuss two models where the $X$ couples to either a purely right-handed or a purely left-handed top quark. One of the models is a minimal extension of the SM, while the other model is more complicated. Both models however explain the DM, baryon coincidence puzzle and the interactions which produce the monotop signal at the LHC arise from the sector of the theory which explains the baryon abundance. In the literature, effective theory Lagrangian has been considered where the monotop signal involves interactions that contain both chiralities of the top quarks~\cite{monotops}. Investigating the energy and transverse momentum distributions of the top quark decay products, we show how the top quark chirality can be a useful handle in distinguishing these models at the LHC even after including the SM backgrounds.

The rest of the paper is organized as follows. We discuss models with isospin singlet and doublet $X$ fields in Section~\ref{sect:model}, collider monotop signals from these models in Section~\ref{sect:monotop}, top quark chirality discrimination in Section~\ref{sect:chiral}, and then conclude in Section~\ref{sect:conclusion}.

\section{Models with Explicit Isospin Structure}
\label{sect:model}

In this section, we introduce two minimal extensions of the SM that include color-triplet scalar fields with baryon number violating interactions. The first model (Model 1) includes two iso-singlet color-triplet scalars $X_{1,2}$ with hypercharge $+ 4/3$ and a singlet fermion $N$ with the following Lagrangian, 

\bea 
\label{eq:lagrangianS}
{\cal L}_{1} & = & \lambda_{1}^{\alpha, i} X^*_\alpha N u^c_{i} + \lambda_{2}^{\alpha, i j} {X}_\alpha {d^c_i} d^c_j  + {\rm h.c.}\,  \\
& + & {1 \over 2} m_N NN + m^2_\alpha |X_\alpha|^2 \, . \nonumber
\eea
Here $\alpha$ denotes coupling to different $X_\alpha$, and $i,~j$ are flavor indices (color indices are omitted for simplicity), and we note that $\lambda_2$ is antisymmetric under $i \leftrightarrow j$. 

The second model (Model 2) includes iso-doublet color-triplet scalars $X_{1,2}$ with hypercharge $+1/3$, iso-doublet fermions $Y$ and ${\bar Y}$ with hypercharge $+1$ and $-1$ respectively, and a singlet fermion $N$ with the following Lagrangian     

\bea
\label{eq:lagrangianD}
{\cal L}_{2}  & = & y_1^{\alpha,i} X^*_\alpha {Q}_i N +y^{\alpha,i}_2 X_\alpha \bar{Y} d_i^c \\
&+&y_3^{\alpha,i} X_\alpha {Y} u_i^c +\text{h.c.}\nonumber \\
&+&m_{Y}\bar{Y}Y + {1 \over 2} m_N N N + m_{\alpha}^2 |X_\alpha |^2. \nonumber %\\
%&+& ({\rm kinetic ~ terms}) \,.  \nonumber 
\eea

We note the $Xd^cd^c$ term, which leads to an $s$-channel resonance enhancement of $X$ production at the LHC, is not present in Model 2 due to electroweak charge assignment. Another important difference between these models is that in Model 1 $X$ interacts with only {\it right handed} quarks, while in Model 2 it interacts with the {\it left handed} up-type quarks. We will exploit this feature to distinguish these models at the LHC. 

In Model 1, the exchange of $X$ particles leads to $\Delta B = 2$ processes like double proton decay $p p \rightarrow K^+ K^+$ and neutron $n-{\bar n}$ oscillations. Experimental limits on these processes set stringent constraints on the model parameters (for a detailed discussion, see~\cite{Allahverdi:2013mza}). An interesting aspect of Model 2 is that it does not result in proton decay or $n-{\bar n}$ oscillations. The tightest limits on this model arise from processes like $K^0-{\bar K}^0$ and $B^0-{\bar B}^0$ mixing. 

As pointed out in~\cite{Allahverdi:2013mza}, the fermion $N$ becomes a viable DM candidate in Model 1 provided that $m_p - m_e \leq m_N \leq m_p + m_e$, where $m_p$ and $m_e$ are the proton mass and electron mass respectively. In Model 2, $N$ becomes stable, hence a DM candidate, if $m_N < m_Y$. The measured value of the $Z$ width requires that $m_Y > m_Z/2$. As the current LHC bound on weakly produced doublets is very weak~\cite{Khachatryan:2014mma} and heavily depends on leptonic final states, the iso-doublets $Y$ and ${\bar Y}$ of a few hundred GeV mass can easily evade the current collider searches. 
The mass of the $N$ particle can vary in quite wide range then, however we want to keep it below 100~GeV in order not to make the explanation of the baryon coincidence puzzle difficult. 
For direct comparison between the two models, we consider the case when $m_N \approx 1$ GeV in Model 2 as well\footnote{The SM gauge symmetry allows renormalizable interaction terms in the Lagrangian that include the newly  introduced fields above and leptons such as $H N L$, $X^* L d^c$, $Y L$, and $Y H e^c$. In combination with the terms in Eqs.~(\ref{eq:lagrangianS})~and~(\ref{eq:lagrangianD}), these terms lead to proton decay and/or $N$ decay in the above models. One can forbid these terms by invoking a new continuous or discrete symmetry. A suitable choice, suggested in Ref.~\cite{Allahverdi:2013mza}, is a gauged $U(1)_L$ symmetry that will forbid all of such dangerous terms. Eventually, the breaking of the continuous $U(1)_L$ symmetry can be achieved by providing vacuum expectation value to a Higgs field which possess $U(1)_L$ charge and this way one can generate Majorana mass term for right handed neutrinos, e.g., $\nu^c\nu^c\Delta$. }. 
The prospects for direct and indirect detection of $N$ particle in Model 1 have been discussed in~Ref.~\cite{Allahverdi:2013mza}.

\section{Monotops at the LHC}
\label{sect:monotop}

Phenomenologically the singlet and doublet $X$ scenarios can be differentiated by the single top quark chirality and whether an $s$-channel resonace is present at the LHC:

(1) The $X$ single production is a resonant $s$-channel process in the singlet model.

(2) When coupled to the 3rd generation quarks, the top quark chirality from $X$ decay is opposite between the singlet and doublet cases.

\begin{figure}[h]
\includegraphics[scale=0.6]{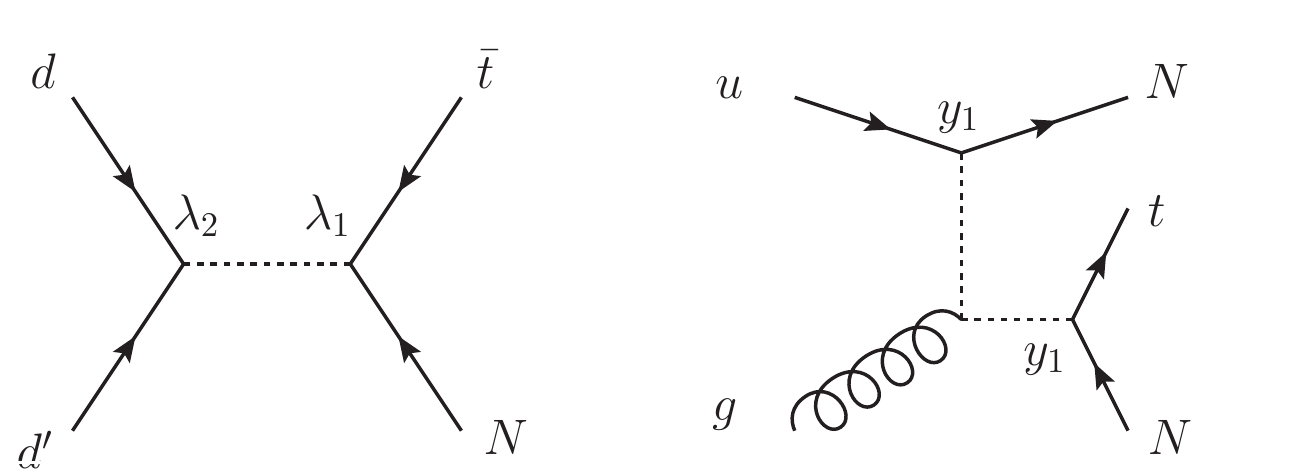}
\caption{Feynman diagrams for single $X$ production at the LHC in the singlet (left) and doublet (right) scenarios.}
\label{fig:singleproductions}
\end{figure}

While $X$ can be singly produced in both models, as shown in Fig.~\ref{fig:singleproductions}, the $X {d}^c d^c$ term in the iso-singlet scenario would allow a resonant $dd' \rightarrow X \rightarrow u N$ monojet process at the LHC, leading to a tight constraint on flavor-blind $\lambda_1$ and $\lambda_2$ couplings as $|\lambda_1 \lambda_2|\le 10^{-2}$. If the top quark couplings $\lambda_{1}^{\alpha, 3}$ is also ${\cal O}(10^{-1})$, a monotop signal will also be expected, and a recent CMS search~\cite{CMS:2016flr} puts a constraint of about 50 fb {\it pre-cut} cross section for a 1TeV resonance's monotop signal. This limit would correspond to $\lambda_1,\lambda_2\sim{\cal O}(10^{-2})$ if the resonance only has the third generation u-quark coupling. In the iso-doublet model, the resonant production is absent and the heavy $X$ in the $t$-channel would yield a smaller production cross section. Generally, if $\lambda$ and $y$ couplings have comparable sizes, the singlet case would be easier to probe at the LHC. Nevertheless, baryogenesis can work for a range of parameter values, the doublet case may also provide a competitive collider signal if $y$ is larger than the singlet coupling. Noted that in the isodoublet model, $X$ couples to $N$ and the doublet $Q$, so a similar diagram to that in Fig.~\ref{fig:singleproductions} can also yield an interesting single $b$ jet + $\met$ final state. We do not study this channel here as the $b$ quark chirality is obscured in its hadronization process.

Alternative channel(s) are certainly also interesting tests of Models 1 and 2. For instance, the production  of the color triplet occurs via the coupling to a pair of down type quarks in Model 1. The decay of the color triplet will involve monojet and/or monotop and two down type quarks (dijet final state). We need the color triplet couplings to down type quarks and an up type quark  plus missing energy to explain baryogenesis. We have studied the dijet, pair of dijet and monojet constraints on the parameter space in Ref.~\cite{Dutta:2014kia}. In this paper, we consider a situation where the up type quark is a top quark and we have monotop final states along with dijet final states. We choose the parameter space $\lambda_2\sim 10^{-2}$ such that both dijet and the recent monotop~\cite{CMS:2016flr} constraints are satisfied for a 1 TeV scalar, and explore the kinematics in the monotop signal events. For Model 2, the production of monotop signal depends only on $y_1$ and if the third generation quarks have larger couplings, monotop can be produced in $X$ decays. In model 2, an interesting case is $y_1\gg y_2,y_3$ that enhances the doublet $X$ decay into left-handed quarks. If we want to produce less than 20fb cross-section in order to satisfy the CMS bound, we only need $y_1<$0.8 for 1 TeV scalar, assuming universal $y_1$ for three quark generations. So we see that monotop signal arises from the interactions responsible for baryogenesis and the parameter space which we use for the monotop signal is not ruled out by any other constraint.   

If monotop signals are present in one or both of the models, 
the chirality of decay products of $X$ will provide a useful tool to distinguish between the two models. Due to a large mass gap between $m_N \approx 1$ GeV and $m_X \sim 1$ TeV, the top quark from $X$ decay gets a significant Lorentz boost and statistically leaves the imprint of its polarization in its decay products. The iso-singlet $X$ decays to a purely right-handed up-type quark, while the iso-doublet $X$ decays to purely left-handed. When $X$ couplings to light quarks are taken into account, 
the $pp \rightarrow t + \met$ process becomes a perfect channel to probe the chiral nature of the $X$ coupling to quarks.

\section{Top chirality as a discriminator}
\label{sect:chiral}

Due to the enhancement from a large top quark Yukawa coupling, the top quark mostly decays into a longitudinal $W$, and the $b$ quark spin aligns with the parent top quark spin in the center-of-mass frame. As $W$ boson only couples to the left-handed current, the direction of $b$ quark momentum would be anti-parallel to the spin and align against the Lorentz boost if the top quark is right-handed. Similarly in the left-handed top quark decay, $b$ quark momentum would be along the Lorentz boost and become more energetic in the lab frame. The top quark polarization can be clearly distinguished with a model-independent observable in the $b$ energy ratio $x=E(b)/E(t)$, which is constructed from the top quark sub-system in the final state:
\be
\zeta \equiv \frac{\Delta N_{+}+\Delta N_{-}}{N_{\text{total}}} 
\label{eq:eta}
\ee
where
\bea
\Delta N_{+}&=&\int_{x_0}^1 \left(\frac{d N}{d x}-\frac{d N^U}{d x}\right) dx \\
 \Delta N_{-}&=&\int_{0}^{x_0} \left(\frac{d N^U}{d x}-\frac{d N}{ dx}\right) dx,\nn
\eea
where $dN/dx$ is a distribution of the bottom quark energy fraction,
$dN^U/dx$ denotes the spectrum from unpolarized top quarks and $\Delta N_{\pm}$ are the deviations above/below the cross-over point $x_0$ at about half the maximum energy fraction, where the pure left/right handed spectra meet. 

In the following discussions about top quark chirality determination, we use a toy model which is based on Model 1 described in section~\ref{sect:model}. This toy model allows to flip the chirality of the produced top quark while leaving the cross section and production mechanism intact. We denote right-handed (RH) model the one producing right-handed polarized top quark. Left-handed (LH) is used to denote the (chirality-flipped) left-handed monotops with the same final state kinematics for mere comparison purposes, not to be confused with Model 2. We do not analyze Model 2 in following sections due to its 100\% left-handed chirality that is the same as the SM single-top background. Model 2 can have a different production process compared to the toy model with left-handed top plus missing energy final states which means that the coupling strengths of the new particles with top quarks are different compared to the toy model. However, the difference in the production process does not change the analysis since our analysis applies cuts on total missing energy, attempts to reconstruct the  top quark and investigate b jet energy distribution. All the selection cuts we have developed can be applied to any scenario of top plus missing energy final state independent of the production process.

\begin{figure}[h]
\includegraphics[scale=0.32]{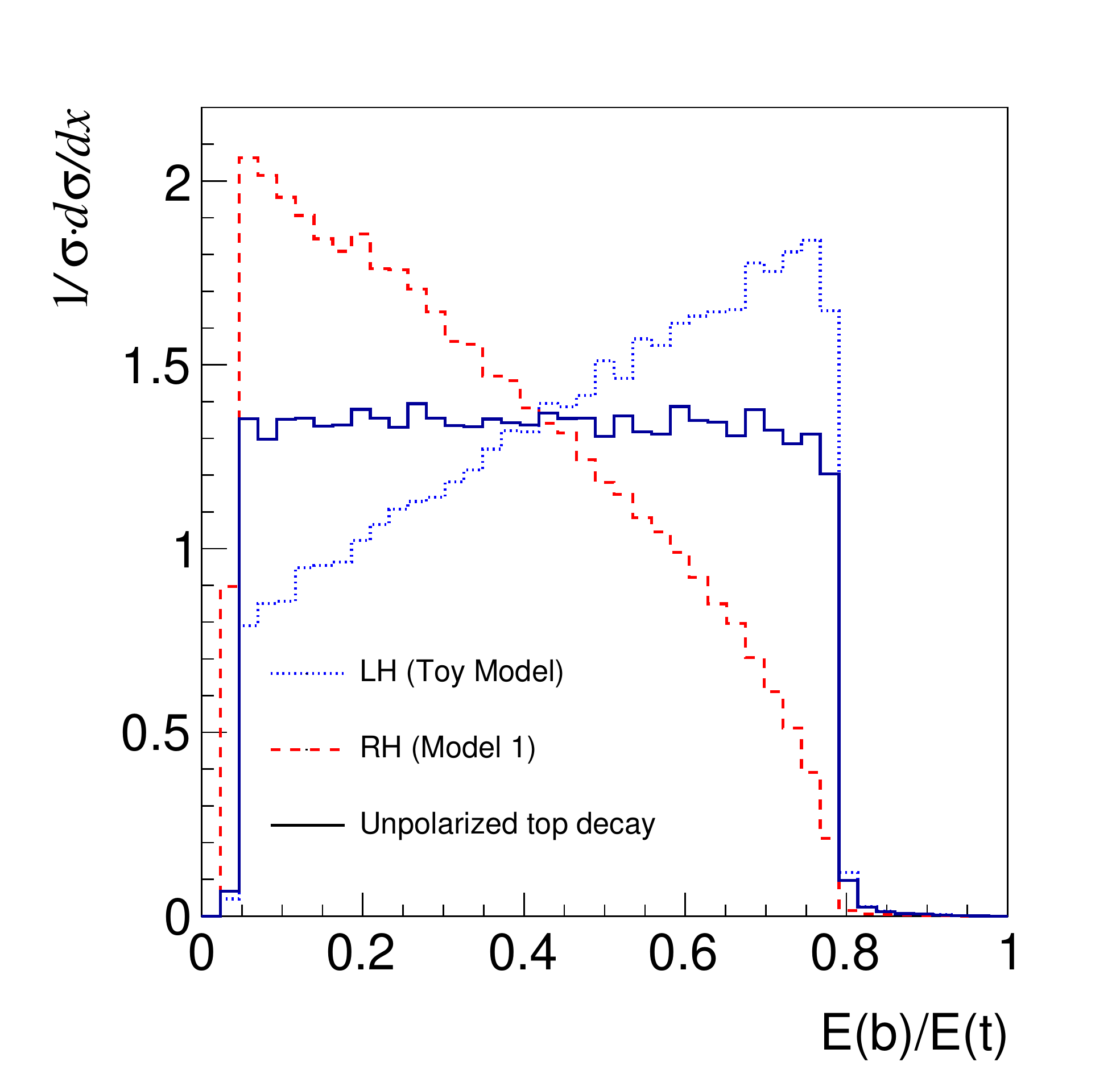}
\caption{Normalized parton level $b$ quark energy fraction from left-handed (dotted) and right-handed (dashed) top quark decays. The flat distribution from unpolarized (solid) top quark decay is shown in black.  $E(t)=500$ GeV in this figure.}
\label{fig:dsdx}
\end{figure}

As shown in Fig.~\ref{fig:dsdx}, a positive(negative) $\eta$ value indicates enhanced left(right)-handed chirality among the top quark sample. The shape of the left and right handed spectra depends on the size of the boost and both distinguish from a flat unpolarized spectrum. The SM single top quarks are mostly left-handed. After subtraction of SM backgrounds, a shape analysis can then determine the top coupling's chirality of a signal, if discovered.

In the analysis of single top quark chirality, the key is a reconstruction of top quark rest frame. In the SM event, the missing energy  is originated by a neutrino in the top leptonic decay. In the monotop event, however, the missing energy is mostly from DM. Unlike the spin correlation in, for instance, $t\bar{t}$ and their decay products, the top quark from our Models 1 and 2 is spin-correlated only with the DM candidate, which is invisible at the LHC. 

Our bottom jet energy fraction approach does not assume any spin correlation between the top quark and other final state particles, thus is independent of the underlying process and can be very useful in cases when not all final state particles are visible and reconstructed.

\begin{figure}[h]
\includegraphics[scale=0.32]{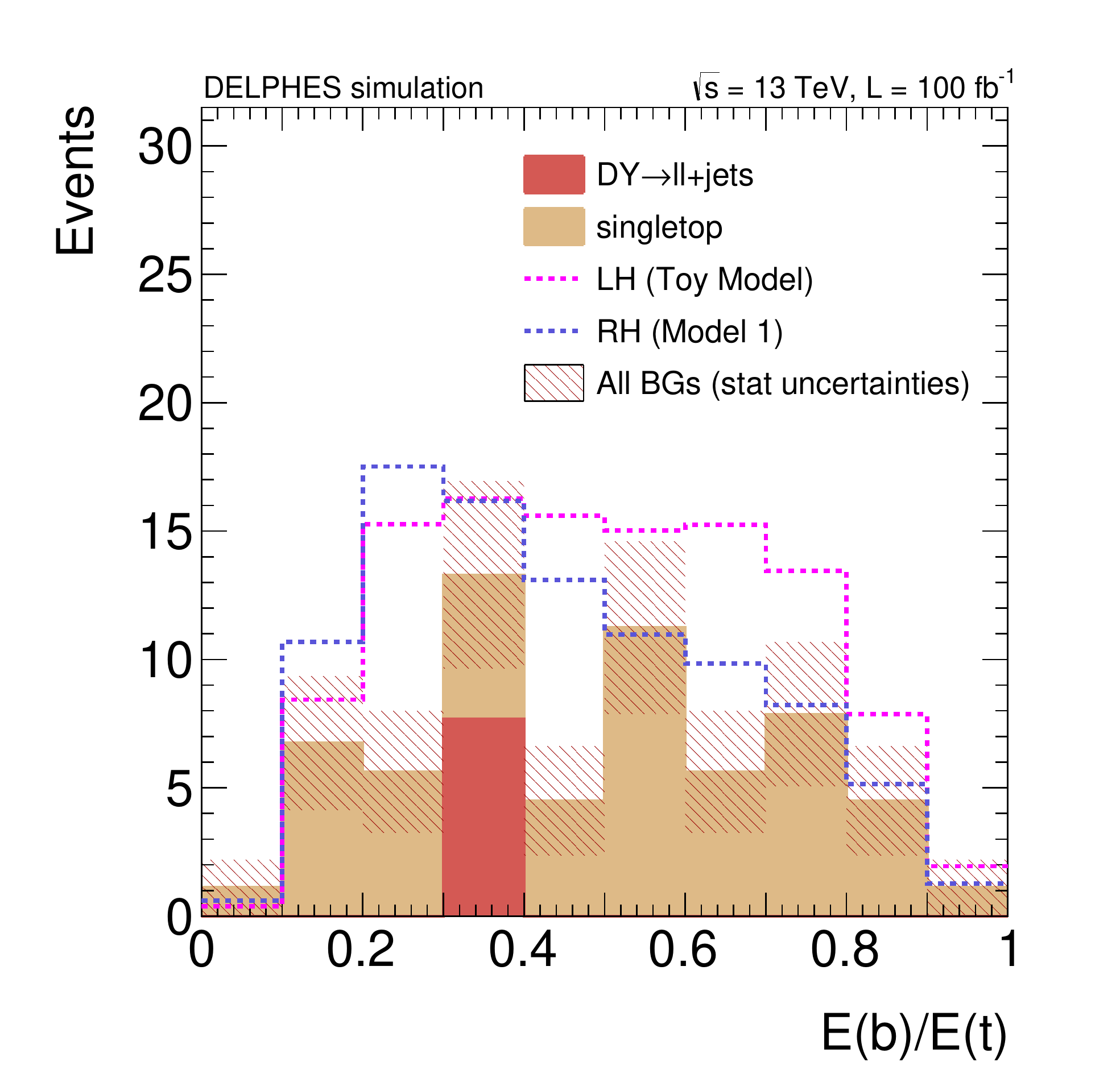}
\caption{Distributions of $b$-jet energy normalized by the top quark energy in hadronic top quark decays for luminosity of 100~fb$^{-1}$ and assumed cross section of 20~fb.}
\label{fig:simres_hadronic}
\end{figure}

The top quark energy can be fully reconstructed for its hadronic decay. When the $b$ jet is successfully tagged, the top quark energy is $E(b)+E(j_1)+E(j_2)$ and $j$ denotes leading non-$b$ jets. The sizable missing energy $\met\sim p_T(t)$ in such events can help reducing SM backgrounds. It is worth notice the SM single-top production is mostly left-handed via the $W$ interaction, while the unpolarized QCD-dominated top quark pair production differs in both top energy/transverse momentum distributions as well as a much more crowded final state. 

The semi-leptonic decay of the top quark also yields different kinematic patterns. Here we include the combination of lepton and $b$-jet energy fractions as a complementary to the fully hadronic channel.  Note the lepton energy ratio to the $b$-jet can also be obtained within the top system, in contrast to  lepton (pair) angular correlations that involves additional final state particles, for instance, in $t\bar{t}$ searches~\cite{Berger:2012an}.

Difficulty may arise from instrumental effects with the detectors as well as kinematic cuts in jet reconstruction. In order to study a feasibility of the proposed method of the top quark chirality determination, we provide the results with a detector simulation. The monotop events are prepared with MadGraph5 v1.5~\cite{bib:madgraph5} for parton-level generation followed by Pythia~8.2~\cite{pythia8} for parton showering and Delphes 3.2~\cite{delphes} for a fast detector simulation. 
In this study we use a default CMS detector card.
The jets are reconstructed with FastJet~\cite{fastjet} package \footnote{The FastJet algorithm used for the jet reconstruction is widely used  and shows very good performance. Moreover, Delphes package was extensively used for various fast simulations predictions and many closure tests which compare the Delphes predictions with full simulation has shown a very good agreement between fast sim and full sim~\cite{delphes}} using anti-$k_{T}$ algorithm for $p_{T} >$ 20 GeV. The efficiency of the $b$-jet tagging is set to be $\sim70\%$ in the barrel part of the detector $(|\eta|<1.2)$ and $\sim60\%$ in the endcaps $(1.2<|\eta|<2.5)$. These numbers correspond to the ones used at the Snowmass workshop~\cite{jetperformanceCMS} (also see~\cite{bjetsperformance})~\footnote{The $b$-tagging efficiencies do not degrade at high transverse momentum as can be seen from the plot:
\url{https://twiki.cern.ch/twiki/pub/CMSPublic/BoostedBTaggingPlots2014/btagperfcomp_Pt700toInf_FatJets_Subjets_StdJets_AK_Hadronic_top.png}. }.

%The fully hadronic events are selected with $p_{T}(b) >$ 60 GeV and at least two jets with $p_{T}(j) >$ 20~GeV. The invariant mass of two light flavor jets is required to be within 20 GeV from the $W$ mass. The semileptonic events are selected with $p_{T}(b) >$ 30 GeV and a lepton with $p_{T}(\ell)>$ 30~GeV. 

The event selection criteria used for the leptonic and hadronic channels are outlined in Tables \ref{tb:2.3} and \ref{tb:2.4}. The criteria were selected in order to reduce the background to 
a minimum level, in order to obtain the best possible signal significance to probe the helicity models under study. Some of the these cuts are reported on the experimental searches by the CMS collaboration, for the leptonic~\cite{Khachatryan:2014uma} and hadronic final states \cite{CMS:2016flr}. The criteria for the $\met$, $m_{T}$ and $m(j, j, b)$ selections, were obtained through an optimization process using the $\frac{S}{\sqrt{S+B}}$ figure of merit for the significance, where $S$ represents the expected signal yield and $B$ the sum of the total background.

 \begin{table}
\begin{center}
\caption {Event selection criteria used for the leptonic channels.}
\begin{tabular}{ l  c }\hline\hline

Criterion & Selection \\
\hline
 $N(b$-jet$)$ & $= 1$ \\
 $p_{T}(b$-jet$)$ & $> 70$ GeV \\
 $|\eta(b$-jet$)|$ & $< 2.5$  \\ 
 $N($non-$b$~jet$)$  & $= 0$  \\
 $p_{T}($non-$b$~jet$)$ & $> 30$ GeV \\
 $|\eta($non-$b$~jet$)|$ & $< 2.5$  \\   
 $N(\ell)$ ($\ell = \mu,e$  )  & $=1$ \\
 $p_{T}(\ell)$ & $> 30$ GeV \\
 $|\eta(\ell)|$ & $< 2.1$  \\ 
 $N(\tau)$ & $= 0$\\
 $p_{T}(W)$ & $> 50$ GeV \\
 $|\Delta \phi (\ell, b$-jet$)|$ & $< 1.7$\\
 Overlaps removal  & $\Delta R > 0.3$\\ 
 $E\!\!\!\!/_{T}$ & $> 100 $ GeV\\
 $m_{T}$ & $> 400 $ GeV \\
\hline
\end{tabular}
\label{tb:2.3}
\end{center}
\end{table}

 \begin{table}
\begin{center}
\caption {Event selection criteria used for the hadronic channels.}
\begin{tabular}{ l  c }\hline\hline
Criterion & Selection \\
\hline
$N(b$-jet$)$ & $= 1$ \\
$p_{T}(b$-jet$)$ & $> 70$ GeV \\
$|\eta(b$-jet$)|$ & $< 2.5$  \\ 
$N($non-$b$~jet$)$  & $\ge 2$  \\
$p_{T}($non-$b$~jet$)$ & $> 30$ GeV \\
$|\eta($non-$b$~jet$)|$ & $< 2.5$  \\   
$N(\ell)$ & $=0$ \\
$E\!\!\!\!/_{T}$ & $> 350	 $ GeV\\
$m(j, j, b)$ & $< 450	 $ GeV\\
\hline
\end{tabular}
\label{tb:2.4}
\end{center}
\end{table} 

\begin{figure}[h]
\includegraphics[scale=0.32]{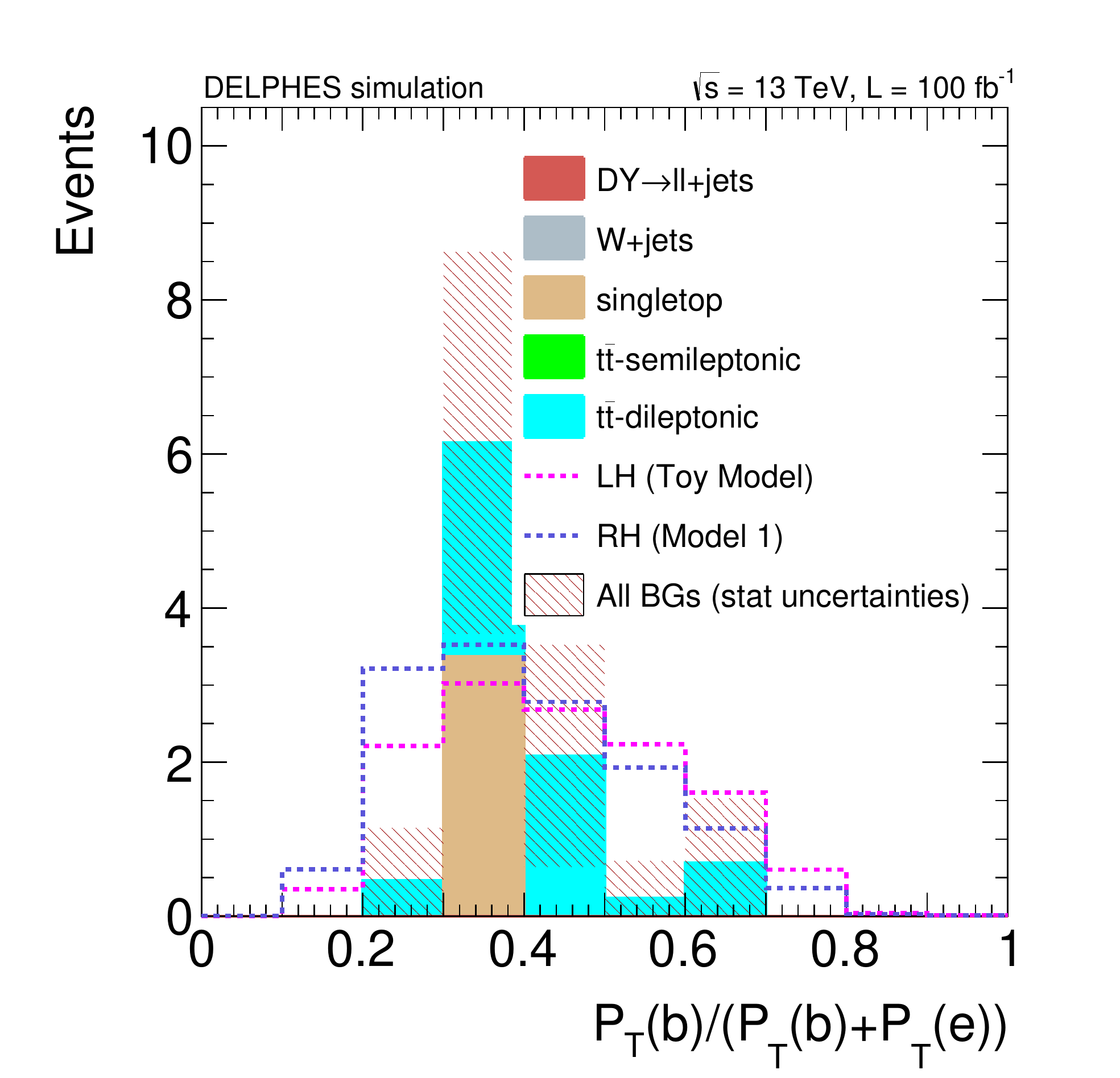}
\caption{Distributions of $b$-jet transverse momentum normalized by the $b+$lepton transverse momentum in $t\to bW\to b e \nu$ top quark decays for luminosity of 100~fb$^{-1}$ and assumed cross section of 20~fb.}
\label{fig:simres_ele}
\end{figure}

\begin{figure}[h]
\includegraphics[scale=0.32]{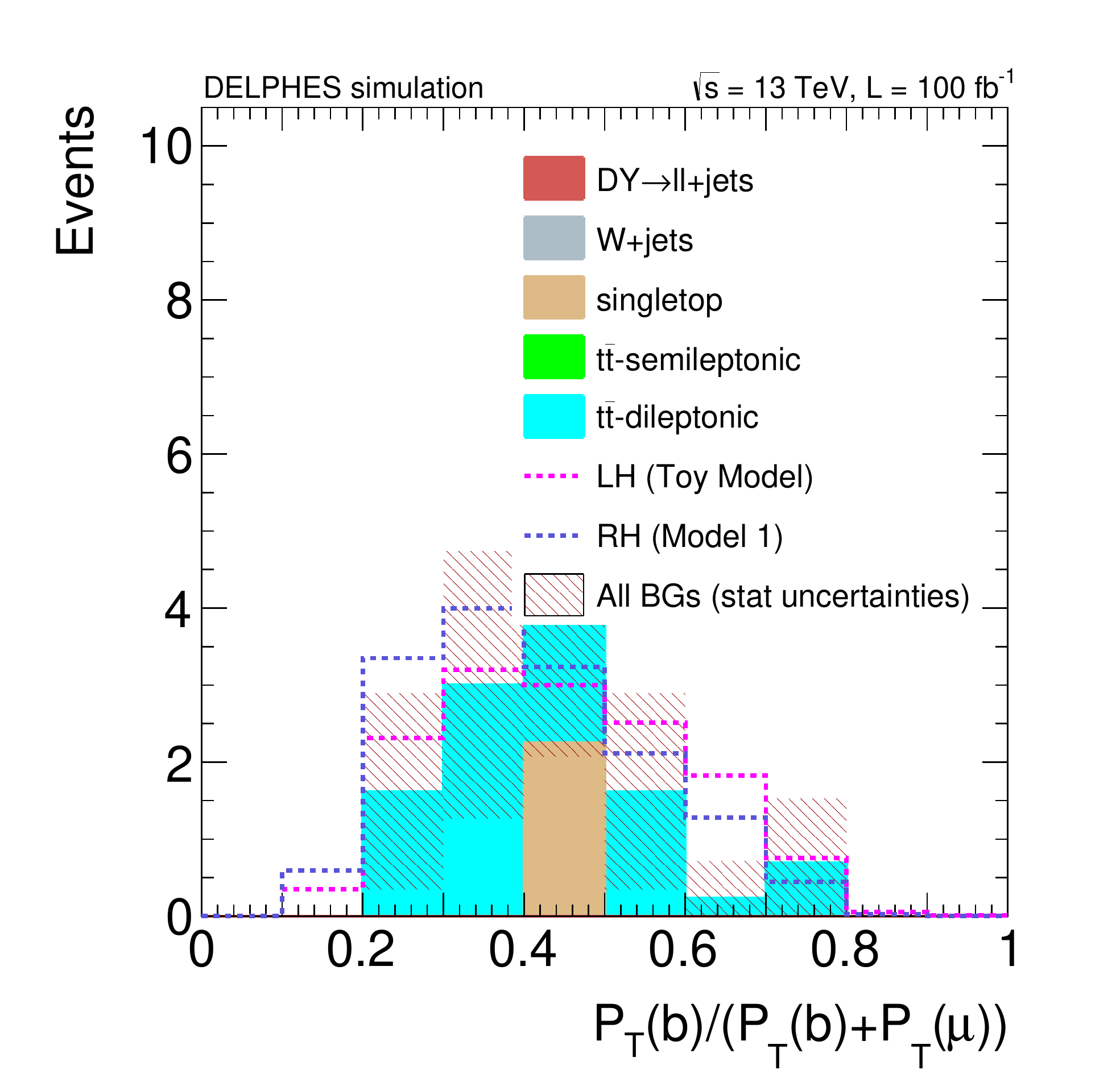}
\caption{Distributions of $b$-jet transverse momentum normalized by the $b+$lepton transverse momentum in $t\to bW\to b \mu \nu$ top quark decays for luminosity of 100~fb$^{-1}$ and assumed cross section of 20~fb.}
\label{fig:simres_mu}
\end{figure}

The results of the simulation are presented in Fig.~\ref{fig:simres_hadronic} for the fully hadronic final state, and in Figs.~\ref{fig:simres_ele} and ~\ref{fig:simres_mu} for the leptonic (electron and muon respectively) cases in which we define the chirality observable as $p_{T}(b) / \left[ p_{T}(b) + p_{T}(l) \right]$. On these distributions, we stack all the contributions from different background processes and the signal contributions are ovelayed. The dashed areas represent combined statistical uncertainties on all the background processes. 

At detector level the $E(b)/E(t)$ spectra shift towards lower values due to smearing and kinematic cuts.  The left-handed case is especially affected as the high $E(b)/E(t)$ range occurs when one jet from $W$ decay locates close to the $b$-jet and may not be correctly reconstructed as a separate jet. Still, the right-hand sample demonstrates a significantly different shape from those in left-handed and unpolarized samples.

Although a proper subtraction of the SM events is expected to be challenging, results show good prospect on the top-quark chirality reconstruction at the LHC experiments. 
We provide a simple cut-and-count sensitivity estimation. We use the number of events with energy (transverse momentum) ratio of bottom and top quarks for hadronic (leptonic) top decays greater than 0.5 (less than 0.5). 
Only statistical uncertainties are accounted for, therefore we expect somewhat deterioration of the results when considering systematic uncertainties. However we note that our results are based on a simple cut-and-count method and should be improved for a proper shape analysis technique, which goes beyond the scope of current paper. 

\begin{figure}[h]
\includegraphics[scale=0.32]{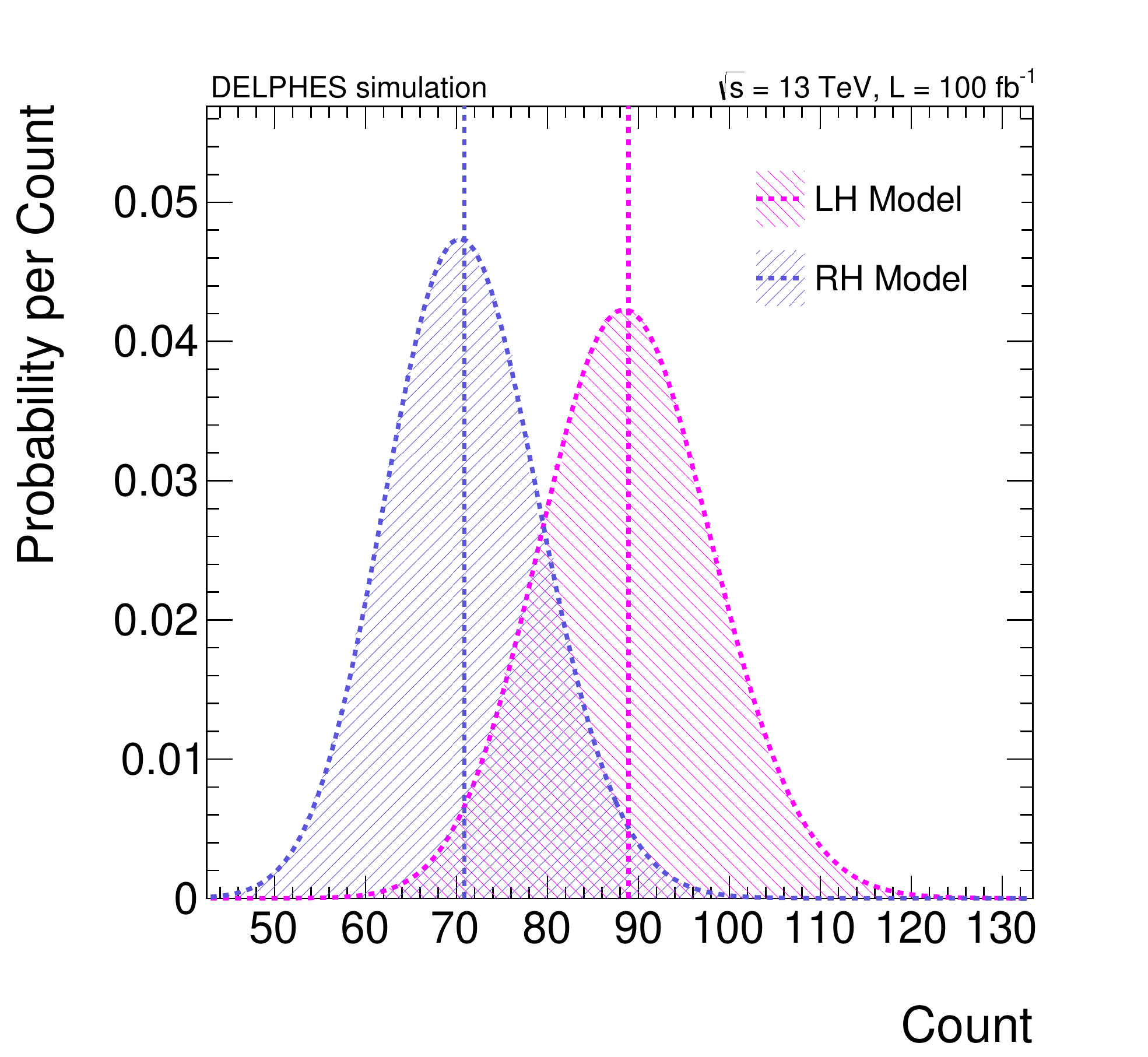}
\caption{Example of test statistic in hadronic top quark decay channel for luminosity of 100~fb$^{-1}$ and assumed signal cross section of 20~fb.}
\label{fig:test_stat_had}
\end{figure}

\begin{figure}[h]
\includegraphics[scale=0.32]{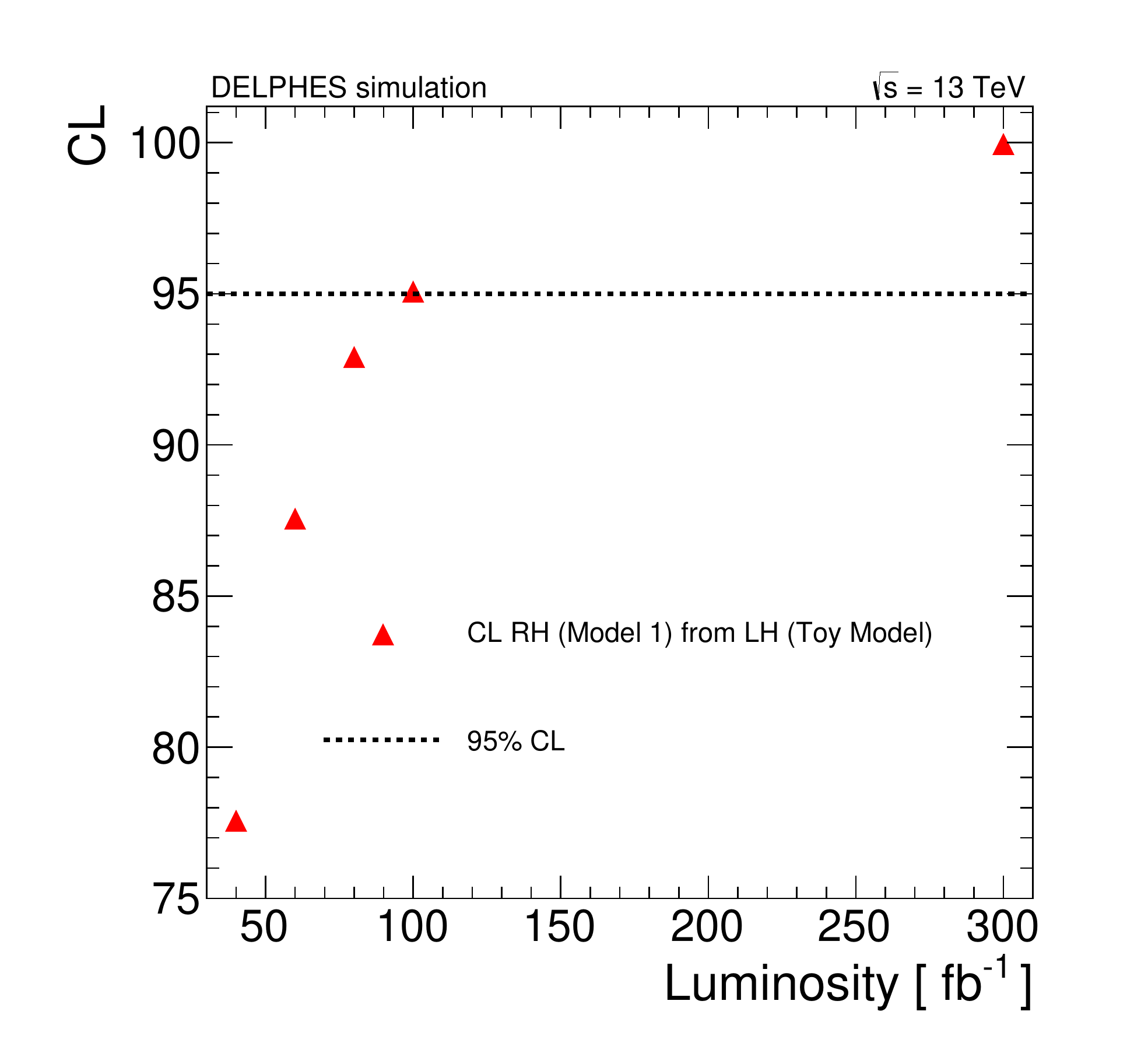}
\caption{Dependency of top quark chirality discriminator separation power on luminosty for hadronic top quark decays. Assumed signal cross section is 20~fb.}
\label{fig:cls}
\end{figure}

We assume the cross section of 20~fb for the signal production, which is below current observed exclusion limits derived from mono- and di-jet search as it was discussed above. Figure~\ref{fig:test_stat_had} shows an example for the test statistics, obtained by generating Poissonian pseudo-experiments around expected yield values for LH and RH signals (on top of background). Figure~\ref{fig:cls} shows the dependency of separation power on luminosity in case of hadronic top quark decays and signal cross section of 20~fb. These figures only include the distinguishment using the $E(b)/E(t)$ in the hadronic channel. The $p_T$ variable of the leptonic channel has a different sensitivity and a similar analysis is performed. The projected luminosities to reach 95\% CL left-right separation for different decay channels of the top quark and different assumed signal production cross sections are summarized in the Table~\ref{tb:lumi}.

\begin{table}[h!]
\begin{center}
  \caption {Projected luminosity (in fb$^{-1}$) needed to reach 95\% CL of separation for different top quark decay modes and signal production cross sections.}
\begin{tabular}{ c c c }\hline\hline
  cross section & 20~fb & 50~fb\\
\hline
decay mode & & \\
$t\to b\mu\nu$ & 1312 & 486 \\
$t\to b e\nu$ & 1842 & 641 \\
$t\to b \ell\nu, \ell=e,\mu$ & 901 & 247 \\
$t\to b j j$ & 97 & 31 \\
\hline
\end{tabular}
\label{tb:lumi}
\end{center}
\end{table} 
\section{Conclusion}
\label{sect:conclusion}

We have discussed minimal extensions of the SM by color-triplet scalars $X_{i}$ that give rise to baryogenesis and a light DM candidate, and lead to highly polarized monotops at the LHC. Polarization of the top quark depends on the electroweak charge assignment of $X$. In the first model, $X$ is an isospin singlet that couples to the right-handed up-type quarks. In the second model, $X$ is an isospin doublet that couples to the left-handed up-type quarks, as well as new color-singlet iso-doublet field(s). This coupling to the up type quark generates the monotop and mono jet  signal. If $X$ couples dominantly to the third generation then we have large branching ratio for monotop final state. Since the new color state $X$ is a vector like color triplet (like right handed up quark) , it can easily be incorporated  under an unifying group like SO(10). 

Unlike the colored superpartners in supersymmetric extensions of the SM, the colored scalars $X$ can be singly produced in both of our models. The models can lead to potentially interesting monotop events where top energy is about half of the $X$ mass, although the monotop production mechanism is different in two models. The large mass of $X$ leads to boosted top quarks whose polarization affects the energy distribution of the decay products. The top quark polarization can therefore serve as a good probe of the isospin of colored scalars and help us differentiate between the two models. The top quark polarization can used to distinguish other monotop models mentioned in Ref.~\cite{monotops} and monotop signal along with the  presence of dijet (Model 1), $W$ (or $Z$) plus jets (Model 2) final states can distinguish the models presented in this paper. 

A detector level simulation shows that the $b$ energy fraction in fully hadronic top quark decays, and the charged lepton momentum fraction in semileptonic decays, can distinguish top quarks of different chiralities. We chose sample signal cross sections that consistent with current searches, and illustrated a projected luminosity for 95\% CL distinguishment between left and right handed monotop samples in Table~\ref{tb:lumi}. SM backgrounds are included in the analyses and selection cuts have been optimized to enhance the signal. 

As chiral couplings between the DM and quarks are often present in beyond SM theories, the search for the top quark chirality can be a very useful probe for establishing such models. Also, while monotop+$\met$ events are often smoking-gun signal for new physics, the spectral analyses of top quark polarization can be readily applied to final state top-quark systems without requiring the top-quark to be spin-correlated with the rest of the final state.

\medskip

{\bf Acknowledgement}

Authors thank D.Rathjens for fruitful discussions. The work of R.A. is supported in part by NSF Grant No.PHY-1417510. The works of B.D. and T.K. are partially supported by DOE Grant DE-FG02-13ER42020. M.D. and Y.G. thank the Mitchell Institute for Fundamental Physics and Astronomy for support. T.K. is also supported in part by Qatar National Research Fund under project NPRP 9-328-1-066. A.F. and M.S. thank the constant and enduring nancial support received for this project from the faculty of science at Universidad de los Andes (Bogot\'a, Colombia), the administrative department of science, technology and innovation of Colombia (COLCIENCIAS). N. K.'s contribution was made possible by the facilities of the Shared Hierarchical Academic Research Computing Network (SHARCNET: www.sharcnet.ca) and Compute/Calcul Canada. We thank the Center for Theoretical Underground Physics and Related Areas (CETUP* 2015) for hospitality and partial support during the completion of this work.

\end{document}